# Econophysical Dynamics of Market-Based Electric Power Distribution Systems

Nicolas Hô and David P. Chassin, *Senior Member, IEEE*

***Abstract*—As energy markets begin clearing at sub-hourly rates, their interaction with load control systems becomes a potentially important consideration. A simple model for the control of thermal systems using market-based power distribution strategies is proposed, with particular attention to the behavior and dynamics of electric building loads and distribution-level power markets. Observations of dynamic behavior of simple numerical model are compared to that of an aggregate continuous model. The analytic solution of the continuous model suggests important deficiencies in each. The continuous model provides very valuable insights into how one might design such load control system and design the power markets they interact with. We also highlight important shortcomings of the continuous model which we believe must be addressed using discrete models.**

***Index Terms*—power system modeling, power system economics, power demand, adaptive control, load shedding, load modeling**

## I. Introduction

The advent of sub-hourly energy markets, such as the New York Independent System Operator's (NYISO) 5-minute clearing time for hub and zone markets, has brought the time-scale of market price fluctuations within the operational time-scale of a significant fraction of load controls and building energy management systems. The commercial and residential sectors account for roughly 2/3 of the total load in the North America and their price response behavior can be a very significant factor in the fluctuation of load at such a time-scale. Recent studies of power price dynamics [1] [2] [3] provide valuable insights into the behavior of power markets as it affects the interplay between supply control and price, but they generally fail to capture that of demand control and price fluctuation at the time-scale that sub-hourly energy markets could operate in. Additionally, a great deal more attention is devoted to understanding and stimulating the response of demand at times when prices are well above average, especially in the context of the mitigation of market power exerted by suppliers [4] [5] [6]. This is of particular concern to those who suggest markets presenting defects must be explicitly mitigated [7] [8]. Unfortunately, relatively little is known about the role of loads when prices are below normal, when they can act then to enhance their response during times of high energy prices.

In exploring the electrodynamic characteristics of electric transmission networks, an important step was made when it was demonstrated that using price signals to affect energy imbalance regulation either by frequency or ACE results in stable control [9]. However, once again more attention has been given to the dynamic interaction between the system as whole and the energy supply resources allocated and dispatched under the influence of market forces than has been given to that of loads. In part, this is caused by the lack of sufficiently detailed models that can couple load behavior in response to system signals such as price over sub-hourly time-scales. But there has historically been a lack of interest in load models working at the minute time-scale because of 1) the traditional belief in the power industry that the load is a boundary condition considered relatively constant within the context of system dynamics, 2) the purpose of regulation is to follow the load rather than include it in the process, and 3) the introduction of such models brings about a degree of computational intensity not previously considered achievable in any practical setting. In the context of this paper, we consider only the latter in the context of two possible remedies, which in time contribute to overcoming the former two. The first is that the availability of high-performance computing to grid operators can be reasonably expected to bring down barrier to load modeling, and the second is that approaches may exist that permit modeling the aggregate dynamic behavior of a large number of discrete load control strategies within the extant computational constraints.

The use of market-based strategies to control loads in buildings is not a new idea. It is based on the widely accepted theory that economic negotiation strategies give rise to an emergent optimal allocation of limited resources, in this case, optimal in the sense of globally cost-minimal or locally profit maximal. Its application to automation systems was initially proposed in the context of allocating computing resources [10] and later extended to building controls [11].

Manuscript received January 9, 2006. This work was funded by U.S. Department of Energy's GridWise Program and conducted by Pacific Northwest National Laboratory, which is operated by Battelle Memorial Institute under Contract DE-ACO6-76RL01830.

David P. Chassin (david.chassin@pnl.gov) and Nicolas Hô (nicolas.ho@pnl.gov) are with Pacific Northwest National Laboratory, Energy Science and Technology Division, Richland, Washington, 99352, USA.

In this study we examine first a numerical model to elucidate the basic behaviors of a market-based electric power distribution system, where demand acts as a full and equal partner with supply in determining price, and where the model of demand control incorporates some of these bidding and response strategies. The model is then generalized into an analytical model that allows us to examine the influence of specific parameters on the total cost of operating the system. A discussion on the necessity of tailoring the bidding strategy to the market dynamics follows.

## II. Iterative model

This model considers the condition in which we control the heating of a building. Of course, the results apply for a model of air conditioning, water-heating, etc. Auxiliary sources of heat not participating in the bidding strategy, such as internal gains and solar gains, are not considered. A simple model for simulating the bidding strategy consists of two iterative equations; one to determine the bid price, the other to determine the temperature of the building. At every time step, the bid price is compared to a market price, and energy is consumed if the bid is greater than the market price, resulting in an increased temperature in the building. The bidding strategy is defined by a linear dependence on the temperature $T$ with hard cut-offs at $T_{max}$ and $T_{min}$, the boundaries where energy is never consumed and always consumed, respectively. These boundaries ensure that the temperature stays within a predefined range regardless of the market price and thus protect the systems from damage and occupants from discomfort. The definition of the bid price $B$ for the current temperature $T$ is then

$$B(T)=\begin{cases}\infty, & T<T_{\min}\\ P_0-aT, & T_{\min}<T<T_{\max}\\ -\infty, & T>T_{\max}\end{cases} \tag{1}$$

where $P_0$ is the recent average price of power and $a$ is a positive parameter that determines the level of comfort desired (small values of $a$ giving the lowest comfort and large values giving the highest comfort). This relationship is schematically depicted in Fig. 1 and can be used for bidding based on observed temperature conditions as well as for adjusting thermostat setpoints in response to market clearing prices.

The unit quantity of energy $q(t)$ consumed by a single heating unit is then

$$q(t)=\begin{cases}B(t)<P(t): & 0\\ B(t)\geq P(t): & 1\end{cases}, \tag{2}$$

where $P(t)$ is the market price. The energy is constantly consumed until the next clearing of the market, when the value of $q(t)$ is updated again[1]. The temperature change at the time $t$ of a building with a single heating unit is then defined as

$$\frac{dT(t)}{dt}=\alpha q(t)-R\left(T(t)-T_{out}\right), \tag{3}$$

where $T$ is offset by the setpoint temperature so that $T=0$ when $P(t)=P_0$, $\alpha$ is the heating rate of the building, and $R$ the natural heat loss rate of the building when the outside temperature is $T_{out}$ (considered constant in this model and also offset by the setpoint). Equations (1-3) are iteratively evaluated for every clearing of the market, taken at 1/10 hour intervals. The cost $C(t)$ of heating the building up to the time $t$ is defined as

$$C(t)=\int_0^t P(x)Q(x)dx\,. \tag{4}$$

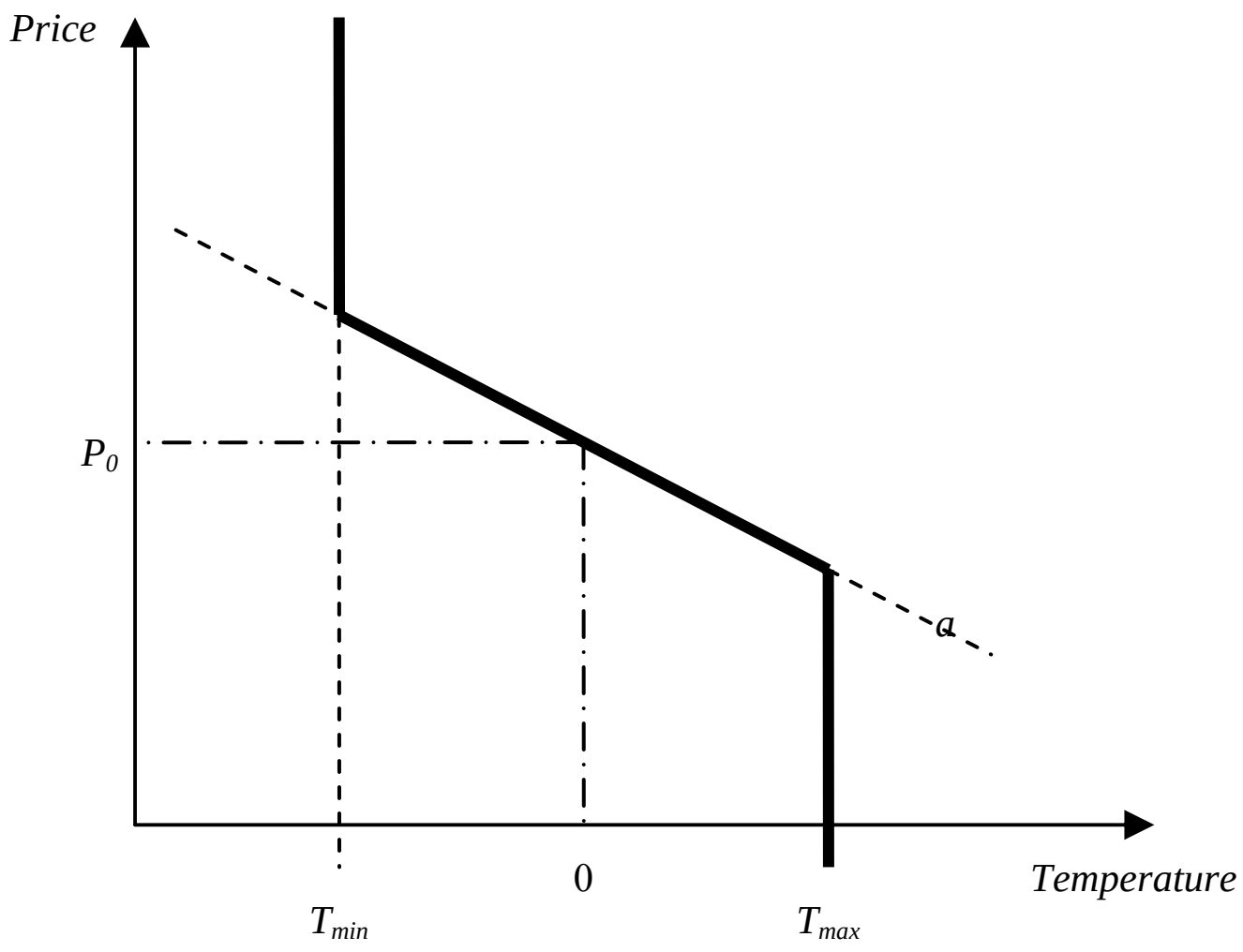

Fig. 1: Relationship in thermostatic loads between temperature deviation from desired indoor temperature and energy price. The slope $a$ and the values of $T_{MIN}$ and $T_{MAX}$ are controlled by the consumer and determine both the bidding strategy and the sensitivity of the load to energy price. Larger values of $a$ improve comfort but reduce economy. Smaller values of $a$ with larger values of $T_{MIN}$ and $T_{MAX}$ result in greater demand elasticity.

The last parameter that needs to be defined is the market price. To compare the behavior of a market-based system a fluctuation market price is compared to a system, operating with a constant price by calculating the cost of operating each. In general, the market price is determined by the collective effect of all the bids and thus, the market price can exhibit very complex behavior, influencing the total cost in unexpected ways. In this, case we are considering the behavior of a single bidder, presumably in the presence of many other bidders, and that lone bidder does not significantly influence the behavior of the market. Consequently we will take the market price $P$ as an exogenous input, which is independent of the bid, $B$.

[1] This is true only in this idealized model. In real systems, $q(t)$ will also change when $T$ goes outside the allowed range of [$T_{min}$, $T_{max}$] and the setpoint hysteresis is reached. For our purposes, it is convenient to not consider this situation.

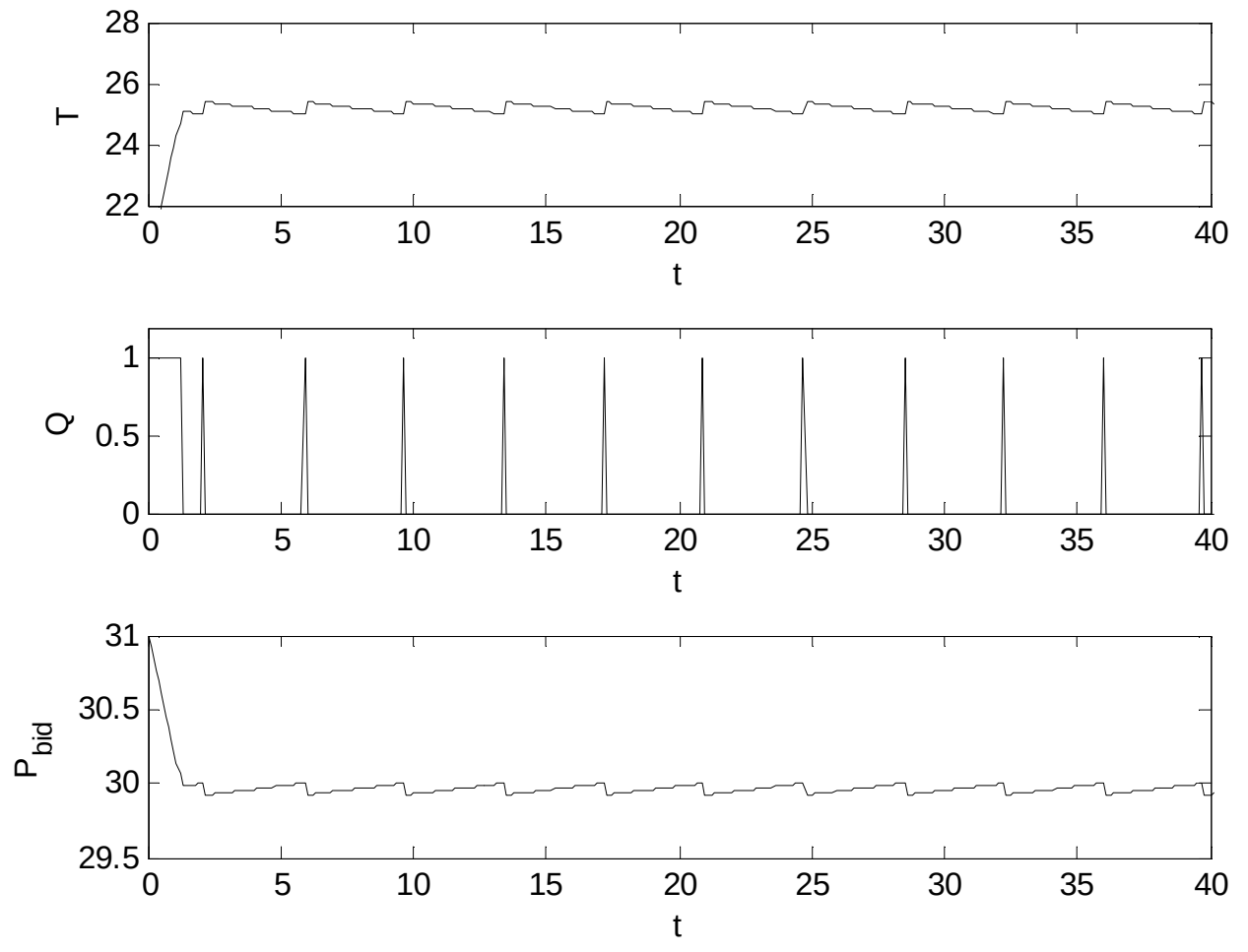

Fig. 2: Temperature, quantity and bid price evolution when the price is constant. The key parameters are $a = 0.2$, $P_0 = 30$, $T_0 = 25$, $R = 0.007$, $T_{min} = 20$, $T_{max} = 30$. The constant price $P = 30$ is indicated by the dashed line.

It is most instructive to study the effect of the specific effect of price fluctuations compared to a constant price, so the market price is taken to be a sine function. The amplitude and frequency of the sine function can be changed with respect to other parameters to illustrate the major contributions of the fluctuations on cost. Certainly a single sine function is far from representing a real market price evolution. Nonetheless, most market time-series can be represented as sums of sine functions, such as a Fourier series, and in that sense, our conclusions retain generality.

The time-series behavior of the key temperature, load, and bid variables when the price is constant is shown in Fig. 2. The temperature always converges to the desired setpoint; with the bid price always higher or lower than the market price depending on temperature, as appropriate. When equilibrium is reached, the temperature undergoes periodic fluctuations in a manner similar to a standard thermostats. When the temperature is high enough, the bid price is lower than the market price and no energy is consumed. This remains so until the temperature is low enough to generate a bid higher or equal to the market price, at which point the temperature increases and the cycle starts again. This behavior is typical of a standard thermostat that is not price sensitive, but also occurs when the clearing price has a zero variance over time, and can be further force by the customer by setting the thermostat's parameter $a$ to a very large number (indicating that maximum comfort is desired).

Naturally, the time-series behavior of the system is not as simple when the price is fluctuating and the effect on total cost can be very different depending on the frequency of the market price fluctuations relative to the time-constant of the building's heat loss. The market price is defined as $P(t) = A\sin(\omega t)$; the period of the market price, $t_P = 2\pi/\omega$; and the relaxation time of the temperature of the building, $t_R = 1/R$ (defined as the time needed to lower the temperature to $1/e$ of its initial value). Fig. 3 shows the cumulative cost for four different market conditions. First, the constant price is shown, where the cost is constantly accumulated, as expected. Second, the market is such that $t_P > t_R$. Third, $t_P = t_R$, where large temperature oscillations occur since the heat loss occur on the same time-scale as the market price evolution. In this case the bid price stays low for a long time before it is high enough to buy energy, at which point a large quantity of energy must be consumed to heat the building to a suitable level. Fourth, $t_P < t_R$, where very small temperature oscillations are present. The market price fluctuates fast enough so that it is always possible to buy energy at low prices and yet keeping the temperature at a desired level, thus resulting in generally low bid prices.

The effect on accumulated cost of the different markets is also represented in Fig. 3. All scenarios are compared to the constant price market. It is clear the case where $t_P < t_R$ offers the largest consistent cost savings, the highest comfort, and that these savings grow linearly in time. That market allows buildings to store heat so that the energy can be bought at only its lowest price. The other two fluctuating price scenarios do allow cost savings, but largely because the temperature is allowed to go down considerably when the price is high. This creates large temperature fluctuations that are undesirable. To the contrary, the case where $t_p < t_R$ shows very fast temperature fluctuations to take advantage of the fast fluctuations of the market price. This is the essence of that strategy: efficient storing of heat that allows the high price peaks to be ignored. This will be impossible if the heat loss of the building occurs rapidly compared to the market price fluctuations. The control by design of these parameters is thus paramount to successful implementation of a market-based bidding strategy.

## III. Continuous model

We now consider a continuous model for the aggregate power $Q(t)$, which will allow the analytical study of the interplay of the different parameters. We recognize two important simplifying assumptions here that are not true in reality, but that we must accept to gain important insights. First, we are assuming that prices change continuously, which in reality they do not. This is an assumption that has been made previously in important models, such as the Black-Scholes model [12] for option pricing without significantly detracting from the insights gained, or indeed from the useful significance of the result. Second we also assume that power consumption varies continuously, which at the level of end-use loads is certainly not the case either. Nevertheless, while this model may deviate in important ways from reality at the level of individual loads, it is expected to provide a relatively accurate depiction of the system behavior in the aggregate.

The two equations of interest are now

$$\begin{aligned} \frac{dQ}{dt} &= \gamma\left(P_{bid} - P(t)\right) = P_0 - P(t) - aT(t) \\ \frac{dT}{dt} &= bQ(t) - r\left(T(t) - T_{out}\right) \end{aligned} , \quad (5)$$

where the parameter $\gamma$ is introduced to convert the difference between the bid price and the market price into a change of power, absorbing the constants of the other variables on the right side of the first equation in (5). The parameter $b$ describes the heat gain rate of the building when energy is consumed, and $r$, the heat loss rate of the building to the outside. All these parameters are positive. The market price $P(t)$ will have the form

$$P(t) = P_0 + A\sin(\omega t) \ , \tag{6}$$

and two cases will be considered; one where the price is constant ($A$=0) and one where the price fluctuates ($A \neq 0$). The cost difference between these two cases will then be determined.

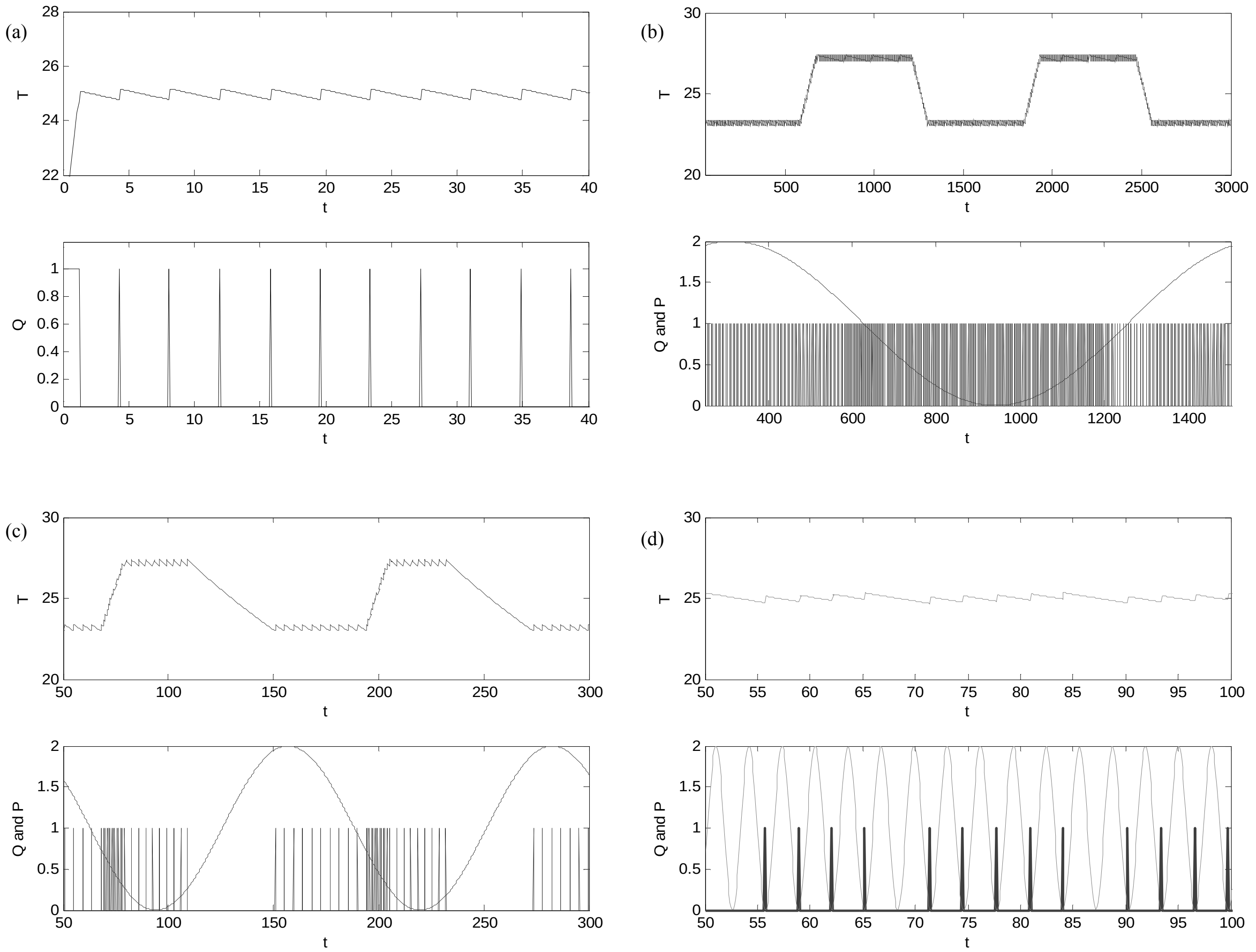

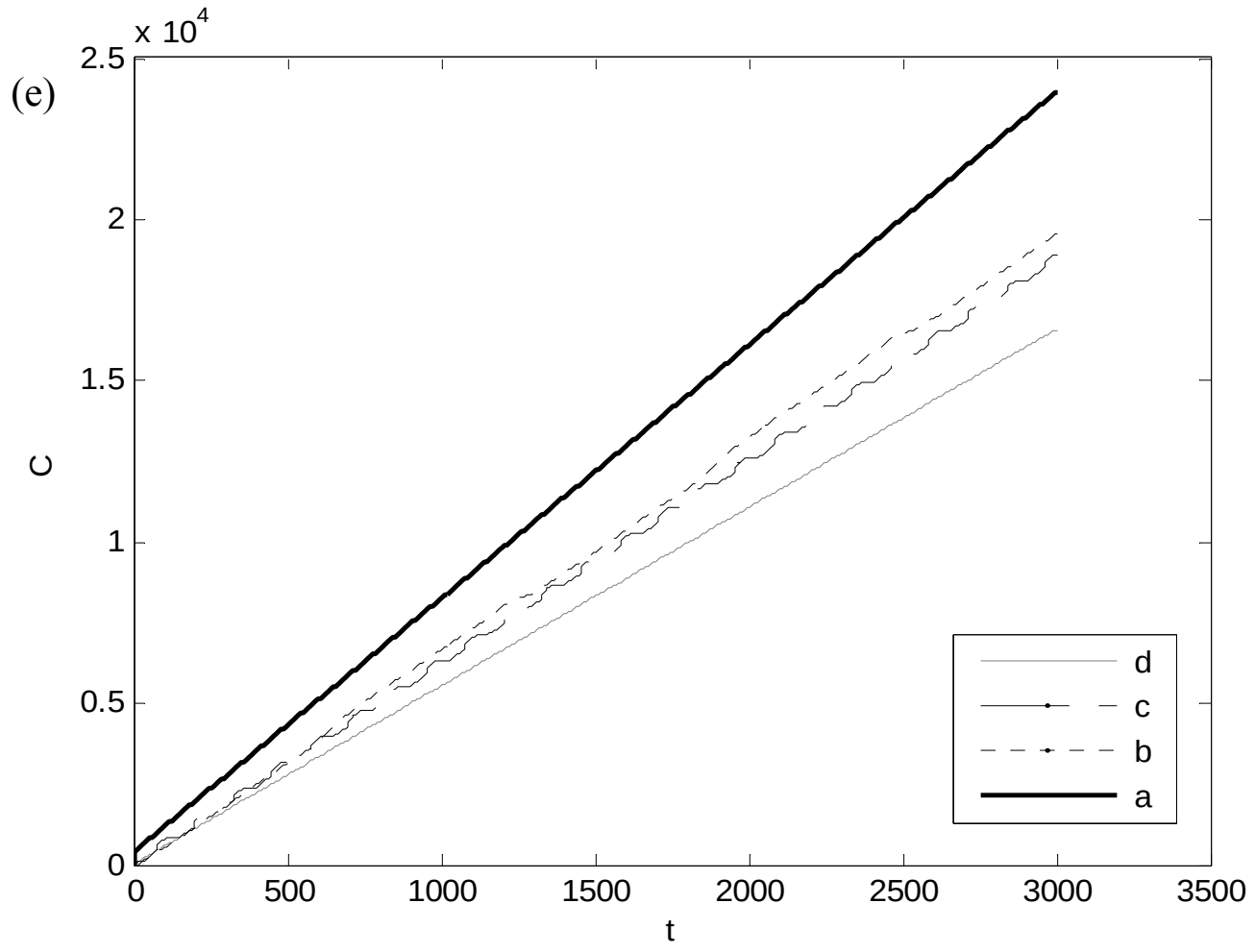

Fig. 3: Effect of $\omega$ and $R$ on system dynamics and cost. $R$=0.007 in all simulations, $a$ = 0.2, $P_0$ varies the average of $T$ at 25, $T_{min}$ = 23 and $T_{max}$ = 27. In graphs (a)-(d), the top plot shows the temperature and the bottom plot shows normalized energy and price. In (a) price $P(t)$ = 25. In (b) price $P(t) = \sin(0.005t)$ s.t. $t_P > t_R$. In (c) price $P(t) = \sin(0.05t)$ s.t. $t_P \approx t_R$. In (d) price $P(t) = \sin(0.2t)$, $t_P < t_R$. In (e) are the cumulative cost for the conditions of (a)-(d).

First, the case where $P(t) = P_0$ is solved. Equations (5) form a system of non-homogeneous differential equations that can be solved by the method of undetermined coefficients[13].

$$Q(t) = -\frac{c_1 a}{\lambda_1} e^{\lambda_1 t} - \frac{c_2 a}{\lambda_2} e^{\lambda_2 t} - \frac{r}{b} T_{out}$$
$$T(t) = c_1 e^{\lambda_1 t} + c_2 e^{\lambda_2 t} \quad (7)$$

with

$$\lambda_1 = \frac{1}{2}\left(-r + \sqrt{r^2 - 4ab}\right), \quad \lambda_2 = -\frac{1}{2}\left(r + \sqrt{r^2 - 4ab}\right) \quad (8)$$

and the constants $c_1$ and $c_2$ can be found with initial conditions. The solution implies that after a transient period dominated by the exponential term, the system will settle to a constant steady state, dominated by the last term on the right of equations (7). The temperature will settle to $T = 0$.

For $P(t)=P_0+A\sin(\omega t)$, the oscillation of the price will drive both $Q(t)$ and $T(t)$ into oscillation at the same frequency. The solution of the system is

$$Q(t) = -\frac{d_1 a}{\lambda_1} e^{\lambda_1 t} - \frac{d_2 a}{\lambda_2} e^{\lambda_2 t} + \sqrt{u_1^2 + v_1^2}\sin(\omega t + \delta) - \frac{r}{b} T_{out}$$
$$T(t) = d_1 e^{\lambda_1 t} + d_2 e^{\lambda_2 t} + \sqrt{u_2^2 + v_2^2}\sin(\omega t + \theta) \quad (9)$$

with

$$u_1 = \frac{A\omega\left(r^2 + \omega^2 - ab\right)}{a^2b^2 - 2ab\omega^2 + r^2\omega^2 + \omega^4}$$
$$u_2 = \frac{bAr\omega}{a^2b^2 - 2ab\omega^2 + r^2\omega^2 + \omega^4}$$
$$v_1 = \frac{-abAr}{a^2b^2 - 2ab\omega^2 + r^2\omega^2 + \omega^4}$$
$$v_2 = \frac{bA\left(\omega^2 - ab\right)}{a^2b^2 - 2ab\omega^2 + r^2\omega^2 + \omega^4} \quad (10)$$
$$\delta = \tan^{-1}\left(\frac{u_1}{v_1}\right)$$
$$\theta = \tan^{-1}\left(\frac{u_2}{v_2}\right)$$

and $d_1$ and $d_2$ can be found from initial conditions. Both $Q(t)$ and $T(t)$ converge to the same value as in the case where $P(t) = P_0$, but there is now also an oscillation around that value produced by the sine functions. The oscillation of $Q(t)$ and $T(t)$ occurs at the same frequency as $P(t)$, but with phase shifts, respectively $\delta$ and $\theta$, determined by the different parameter values. Note that in the steady state, the same amount of energy is consumed on average as in the case $P(t) = P_0$. It is in the cost that we will observe a benefit to the control strategy.

The cost, defined in equation (4), can be calculated for both case previously shown. The comparison of the total costs of these models allows determining if cost savings are indeed possible and the value of the parameters to maximize the savings. The cost difference is given by

$$\Delta C(t) = C_{fluct}(t) - C_{const}(t)$$
$$= A\sqrt{u_1^2 + v_1^2}\left(\frac{t}{2}\cos(\delta) - \frac{\sin(2\omega t + \delta)}{4\omega} + \frac{\sin(\delta)}{4\omega}\right), \quad (11)$$

Of all the terms in the last equation from equations (11), only one really matters. The coefficient $A(u_1^2 + v_1^2)^{1/2}$ is always positive, the first term $t\cos(\delta)/2$ increases or decreases linearly depending on the value of $\delta$, the second term $-\sin(2\omega t+\delta)/4\omega$ has a mean of zero and the last term $\sin(\delta)/4\omega$ is constant. The main contribution to the steady state cost savings then comes from the cosine term. If it is negative, the bidding strategy will lead to cost savings that will grow in time. The cosine function is negative for

$$-\pi \le \delta < -\pi/2 \text{ and } \pi/2 \le \delta < \pi \quad (12)$$

on the $[-\pi, \pi]$ interval. The value of $\delta$ is given by

$$\delta = \tan^{-1}\left[\frac{\omega\left(r^2 + \omega^2 - ab\right)}{-abr}\right], \quad (13)$$

and from the definition of the arctangent, $-abr < 0$ is necessary to ensure that $\delta$ will satisfy equation (12). Cost savings are thus achieved by using the control strategy and the cost savings will be maximized for $\delta = \pm\pi$, which will occur for

$$\omega = \sqrt{ab - r^2} \text{ and } ab >> \frac{\omega\left(r^2 + \omega^2\right)}{r + \omega}. \quad (14)$$

The product $ab$ compared to $\omega$ and $r$ is thus of major importance. As can be seen from the last results, increasing the oscillations frequency of price $\omega$ is not sufficient to lower the cost, contrary to the iterative model presented earlier. In the case of the continuous model, it is not possible to skip (i.e., not respond to) a price period. The price drives all variables and they must oscillate at the same frequency. It is the phase shift between the price and the required energy quantity that creates the cost savings; by storing heat in the building when the price is high, energy can be bought at its lowest price. The correct determination of the parameters of the system will enable the larger cost savings.

## IV. Discussion

It is very important to keep in mind that the continuous model does have serious limitations when trying to describe the discrete behavior of individual loads, but the insights gained are nonetheless important. So it is with this caveat that we consider the conclusions we can draw from these results.

The main conclusion of the result is that the implementation of a market-based electrical system will yield different cost-savings depending on the relative rate of the market price changes and the intrinsic parameters of the buildings included in the system. However simple they might be, they highlight the importance of carefully studying the market price evolution in terms of the high frequency it presents to loads and the intrinsic parameters of the buildings (insulation, heat capacity, equipment size, etc.) before choosing the bidding strategy for increasing cost savings.

All the results presented in this report deal with sinusoidal price evolutions. This is the easiest comparison to a constant price system, but all the results can be easily generalized to deal with more than one oscillating period. The basic results stay the same; the relationship between the period of price fluctuations, the intrinsic parameters of the building and the bidding strategy will dictate the cost-efficiency of the market-based system. As long as there are price fluctuations, steady-state cost savings are possible. However, the presented results all considered that the system reaches a steady state. Abrupt changes in market behavior, such as abnormal spiking known to occur, lead to significant transient cost differences that can be either negative or positive, depending on their timing. These events will play an important role in the overall performance of the bidding strategy and further studies are required to quantify their effect in relation to the design of the building energy management systems.

It is also important to note that the inclusion of an effective price forecaster in the bidding strategy could be used to further increase the cost savings. If the occurrence of these events could be foreseen, probably even with relative precision, it would be possible to take advantage of them to either store heat in advance if the price will go up or wait until they occur to buy more energy if the price goes down. However, the efficient market hypothesis suggests that the benefit of forecasters diminishes as they become more widely used. It seems more likely that an efficient load forecaster would be beneficial, and would suffer less from the effect of broad application in power markets.

Clearly, the speed at which the market iterates sets an upper limit on the highest frequency that can drive the loads. The rate at which the bidding system will be updated is therefore very important. Machines need to be able to bid at a frequency high enough so that they can take advantage of fast price fluctuations. Knowing the parameters of the building and system, that rate can be approximated from the models shown before. Rates as fast as 5 minutes should be possible, as evidenced by the operation of the NYISO hub and zone markets, which updates its prices at this rate.

## V. Conclusions

We have shown that a simple control strategy for building energy management systems can result in an overall cost reduction when responding to real-time power prices. The key parameters that determine the savings are the frequency at which the market is cleared, the building's heating and cooling time-constants, and the willingness of the building occupants to accept temperature fluctuations. We have quantified the conditions for which cost-savings can be obtained and observed that these conditions are possible for today's buildings if suitable real-time power markets were available.

## VII. Biographies

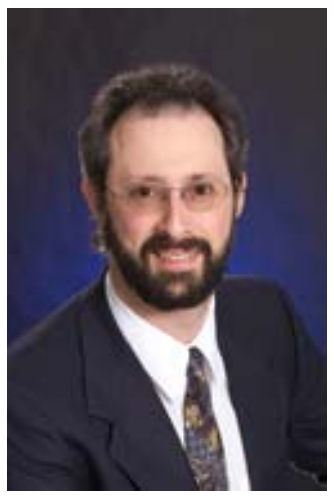

**David P. Chassin** (M '03, SM '05) received his BS of Building Science from Rensselaer Polytechnic Institute in Troy, New York. He is staff scientist with the Energy Science & Technology Division at Pacific Northwest National Laboratory were he has worked since 1992. He was Vice-President of Development for Image Systems Technology from 1987 to 1992, where he pioneered a hybrid raster/vector computer aided design (CAD) technology called CAD Overlay™. He has experience in the development of building energy simulation and diagnostic systems, leading the development of Softdesk Energy and DOE's Whole Building Diagnostician. He has served on the International Alliance for Interoperability's Technical Advisory Group, and chaired the Codes and Standards Group. His recent research focuses on emerging theories of complexity as they relate to high-performance simulation and modeling in building controls and power systems.

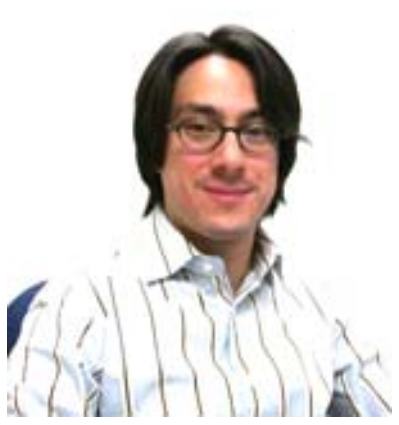

**Nicolas Hô** received his Ph.D. in physics from Université Laval, Quebec, Canada in 2004. He is a scientist with the National Security Directorate at Pacific Northwest National Laboratory since 2005. He works in the fields of optics and photonics with specific interest in integrated optical components for bio-chemical sensing. He also works on theoretical aspects of the dynamics of financial markets and market-based control systems.